\documentclass[10pt,twocolumn]{IEEEtran} 

\usepackage{amsmath,url}
\usepackage{epsfig,amssymb,amsbsy,verbatim,array,enumerate}
\usepackage{pstricks,psfrag,theorem,cite,paralist,subfigure}

\let\intern=\iftrue

\def\figref#1{Fig.\,\ref{#1}}%
\def\E{\mathbb{E}}
\def\P{\mathbb{P}}
\def\R{\mathbb{R}}

\def\ie{{\em i.e.}}
\def\eg{{\em e.g.}}

\def\sir{\mathrm{SIR}}

\def\isr{\mathrm{I{\bar S}R}}

\def\misr{\mathrm{MISR}}

\newtheorem{definition}{Definition}

\newlength{\figwidth}


\begin{document}
\title{The Mean Interference-to-Signal Ratio and its \\Key Role in Cellular and Amorphous Networks} 
\author{Martin Haenggi \\Department of Electrical Engineering\\{University of Notre Dame}
\thanks{Manuscript date \today. The support of the NSF (grants CNS 1016742 and CCF 1216407) is gratefully acknowledged.}
}

\maketitle
\begin{abstract}
We introduce a simple yet powerful and versatile analytical framework to approximate
the SIR distribution in the downlink of cellular systems. It is based on the 
{\em mean interference-to-signal ratio} and yields the horizontal gap (SIR gain) between
the SIR distribution in question and a reference SIR distribution.
As applications, we determine the SIR gain for base station silencing, cooperation,
and lattice deployment over a baseline architecture that is based on a Poisson deployment
of base stations and strongest-base station association. The applications demonstrate that
the proposed approach unifies several recent results and provides a convenient framework
for the analysis and comparison of future network architectures and transmission schemes,
including {\em amorphous networks} where a user is served by multiple base stations and,
consequently, (hard) cell association becomes obsolete.
\end{abstract}
\section{Introduction}
\subsection{Motivation and contribution}
The SIR distribution is a key metric in interference-limited wireless systems.
Due to high capacity demands and limited spectrum, current- and next-generation
cellular systems adopt aggressive frequency reuse schemes, which makes
interference the main performance-limiting factor. 
To overcome coverage and capacity problems due to interference, many
sophisticated transmission schemes, including base station cooperation and
silencing, successive interference cancellation, multi-user MIMO, and multi-tier
architectures have recently been proposed. However, a simple evaluation and
comparison of their effect on the SIR distribution has been elusive.

In this paper, we propose a novel technique that provides tight approximations
of the SIR gain of advanced downlink architectures and cooperation
schemes over a baseline scheme. It is based on the mean
 interference-to-signal ratio (MISR), which is used to quantify
the {\em horizontal gap} between two SIR distributions.
To account for the spatial irregularity of current and future cellular system,
we use point process models  for the positions of the base stations (BSs)
\cite{net:Dhillon12jsac,net:ElSawy13tut}.

\subsection{The horizontal gap in the SIR distribution}
\begin{figure}
\centerline{\epsfig{file=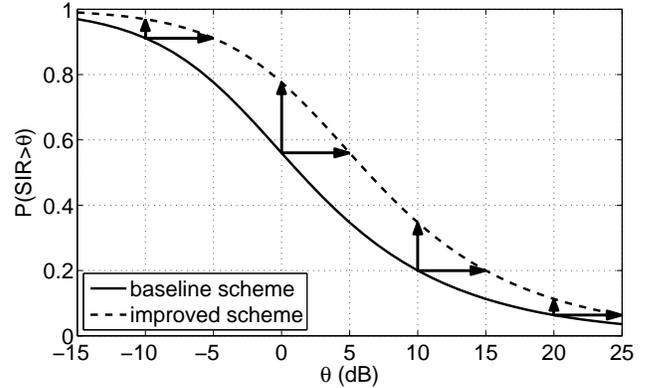,width=\figwidth}}
\caption{Example SIR ccdfs for a baseline and an improved scheme.
The vertical gap between the distributions depends
strongly on the value of $\theta$ where it is evaluated, while the horizontal gap is almost
constant.} 
\label{fig:generic_sir}
\end{figure}

We focus on the complementary cumulative distribution (ccdf)
$\bar F_{\sir}(\theta)\triangleq\P(\sir>\theta)$ of the SIR\footnote{The ccdf is often referred
to as the transmission success probability, while its complement, the cdf, is the
outage probability.}. There are two ways to compare 
SIR distributions, vertically or horizontally, see \figref{fig:generic_sir} for an
illustration. Using the vertical gap,
\ie, the gain in the coverage probability, has several disadvantages:
(1) it depends strongly on the value of $\theta$ where it is evaluated; (2) it is often unclear whether the gain is measured in absolute or relative terms (for example, at -10 dB, the gap is 0.058, or 6.4\%;
at 0 dB, the gap is 0.22, or 39\%, and 
at 20 dB, the gap is 0.05, or 78\%---and sometimes even the absolute gain is expressed in percentages);
(3) the gain also depends heavily on the path loss law and fading models.

In contrast, the horizontal gap (SIR gain)
is often quite insensitive to the probability where it is evaluated and the path loss models.
Formally, the gap between the distributions of $\sir_1$ and $\sir_2$ is defined as
\begin{equation}
G(p)\triangleq\frac{\bar F_{\sir_2}^{-1}(p)}{\bar F_{\sir_1}^{-1}(p)},\quad p\in(0,1),
\label{gp}
\end{equation}
where $\bar F_{\sir}^{-1}$ is the inverse of the ccdf of the $\sir$, and $p$ is the target
success probability.
In \figref{fig:generic_sir}, for example, $G(p)=5 $dB, irrespective of $p$.
In the following, we will illustrate that this behavior is commonly observed.
We also define the asymptotic gain (whenever the limit exists) as
\begin{equation}
G\triangleq G(1)=\lim_{p\to 1} G(p).
\label{gp2}
\end{equation}

\section{The Mean Interference-to-Signal Ratio and its Significance}
\subsection{Definition}
We first define the interference-to-average-signal ratio $\isr$. 
\begin{definition}[$\isr$]
The {\em interference-to-average-signal ratio} $\isr$ is defined as
\[ \isr\triangleq\frac{I}{\E_h(S)} ,\]
where $I$ is the sum power of all interferers and
$\bar S=\E_h(S)$ is the signal power averaged over the fading.
Its mean is denoted by $\misr\triangleq \E(\isr)$.
\end{definition}
The bar over the $S$ in the $\isr$ indicates averaging over the fading.
The $\isr$ is a random variable due to the random positions of the BSs relative to the
typical user. 
For the following discussion, we assume
a power path loss law $\ell(r)=r^{-\alpha}$ with a path loss exponent $\alpha$
and (power) fading with unit mean, \ie, for all fading random variables,
$\E(h)=1$.  We also assume that the desired signal comes from a single BS at 
distance $R$, while the interferers are located at distances $R_k$
 and their transmit powers (relative to the
one of the serving BS) are $P_k$. In this case, the $\isr$ is given by
\[ \isr=R^\alpha \sum_{k\in\mathcal{I}} h_kP_kR_k^{-\alpha}, \]
where $\mathcal{I}$ is the index set of the interferers and $h_k$ denotes the
channel (power) gain.
The mean follows as
\[  \misr\triangleq \E(\isr)=\sum_{k\in\mathcal{I}} P_k\E \left(\frac {R^\alpha}{R_k^\alpha}\right).\]
So the $\misr$ is a function of the {\em distance ratios} $R_k/R$ between the desired and interfering base stations,
scaled by the relative transmit powers.

\subsection{The asymptotic gap for Rayleigh fading}
The SIR distribution can be expressed using the $\isr$ as
\[ F_{\sir}(\theta)=\P(h\bar S<\theta I)=\P(h < \theta\, \isr). \]
For exponential $h$ and $\theta\to 0$,
\[ \P(h< \theta\, \isr\mid \isr )\sim \theta\, \isr ,\]
thus
\[  \P(h< \theta\,\isr)\sim \theta\, \E(\isr) .\]
So $F_{\sir}(\theta)\sim \theta \,\misr$, and
$\bar F_{\sir}^{-1}(p)\sim (1-p)/\misr$, $p\to 1$.
Consequently, the asymptotic gain between two SIR ccdfs 
\eqref{gp2} can be expressed as
\begin{equation}
  G= \frac{\misr_1}{\misr_2},
  \label{asym_gain}
\end{equation}
and if it is finite, we have $\bar F_{\sir_1}(\theta)\sim \bar F_{\sir_2}(\theta/G)$,
$\theta\to 0$.

We will demonstrate in the
next section that this relationship 
provides an accurate approximation
for the gain also at non-vanishing values of $\theta$, \ie, that
$\bar F_{\sir}(\theta)\approx \bar F_{\sir}(\theta/G)$ for all
practical values of $\theta$.

Other types of fading will be discussed in Sec.~\ref{sec:naka}.

\subsection{The HIP model and the baseline MISR}
The homogeneous independent Poisson (HIP) model was first introduced as a model
for cellular networks in \cite{net:Dhillon12jsac}.
\begin{definition}[HIP Model]
A {\em homogeneous independent Poisson} (HIP) model with $n$ tiers 
consists of $n$ independent Poisson point processes (PPPs) $\Phi_k\subset\R^2$
with intensities $\lambda_k$, $k\in[n]$ and power levels $P_k$. 
$\Phi_k$ is the set of locations of the base stations of the $k$-th tier.
\end{definition}
{\em Remarks:}
\begin{itemize}
\item Alternatively, the HIP model can be defined as follows:
Let $\lambda=\sum_{k\in[n]}\lambda_k$.
Starting with a homogeneous Poisson point process (PPP) $\Phi\subset\R^2$ of intensity $\lambda$,
randomly assign each point $x\in\Phi$ to one of the $k$ tiers $\Phi_k$,
where $\P(x\in\Phi_k)=\lambda_k/\lambda$, independently for each $x$.
\item
 Although the HIP model is a model for a heterogeneous network, we call it
{\em homogeneous}, since it is based on homogeneous PPPs, \ie, the base stations form a spatially homogeneous point process.
\item
The HIP model is doubly independent, since it exhibits neither intra-tier nor inter-tier dependence.
This makes it highly tractable but also makes it less accurate in situations where base stations
are deployed in a repulsive fashion (\ie, with a certain minimum distance) \cite{net:Deng14twc} or
where base stations of different tiers are not placed independently \cite{net:Deng14globecom}.
\item Quite remarkably, for the power path loss law with Rayleigh fading and with strongest-BS
association (on average, \ie, not considering small-scale fading),
the SIR distribution for the HIP model does not depend on the number
of tiers $n$, their densities $\lambda_k$, or their power levels $P_k$ \cite{net:Nigam14tcom}.
For $\alpha=4$, the SIR distribution is given by the extremely simple expression
\begin{equation}
 \bar F_{\sir[4]}(\theta)=\frac1{1+\sqrt\theta\arctan\sqrt\theta} .
 \label{arctan}
 \end{equation}
Also, $\E(S)=\infty$ and $\E(\sir)=\infty$ for all values of $\alpha$
due to the proximity of the nearest BS, and  $\E(I)=\infty$ for $\alpha\geq 4$.
\end{itemize}
Due to its tractability, the HIP model with strongest-base station association 
is the perfect candidate for a baseline model against which the gains of other schemes
can be measured. Since the SIR distribution does not depend on the density or number of
tiers, we use a single-tier model in the following to calculate the MISR for the HIP model.

Let $R_k$ be the distance from the typical user to the $k$-th nearest BS.
Its distribution is given by \cite{net:Haenggi05tit}. The
distribution of the distance ratio $\nu_k=R_1/R_k$ is \cite[Lemma 3]{net:Zhang14twc}
\[ F_{\nu_k}(x)=1-(1-x^2)^{k-1} ,\quad x\in [0,1],\]
and the 
$\alpha$-th moments are
\begin{equation}
   \E(\nu_k^\alpha)=\frac{\Gamma(1+\alpha/2)\Gamma(k)}{\Gamma(k+\alpha/2)} .
   \label{nu_alpha}
 \end{equation}
For equal powers $P_k\equiv 1$, the MISR follows as\footnote{%
The parameter $\kappa^{\rm PPP}$ calculated as the limit
$\lim_{\theta\to 0} F_{\sir}(\theta)/\theta$ in \cite[Cor.~1]{net:Guo14tcom} is identical to the MISR for the PPP.}
\begin{equation}
 \E(\isr)=\sum_{k=2}^\infty \frac{\Gamma(1+\alpha/2)\Gamma(k)}{\Gamma(k+\alpha/2)}=\frac{2}{\alpha-2},\quad\alpha>2.
 \label{misr}
\end{equation}
For $\alpha=4$, $\misr=1$, which implies $F_{\sir}(\theta)\sim \theta$,
$\theta\to 0$.

\section{Applications}
\subsection{Base station silencing}
We consider the (single-tier) HIP model and
let $\isr^{(!n)}$ denote the $\isr$ if the $n$ strongest interfering BSs (on average) are silenced,
and all BSs transmit at the same power.

If the nearest interfering BS is silenced, the MISR is obtained by subtracting
$\E(\nu_2^\alpha)$ from \eqref{misr}, which yields
\[ \E(\isr^{(!1)})=\frac{2}{\alpha-2}-\frac{2}{\alpha+2}  =\frac{8}{\alpha^2-4} .\]

For general $n$,
\[ \E(\isr^{(!n)})=\frac{2\Gamma(1+\alpha/2)}{\alpha-2} \,\frac{\Gamma(n+2)}{\Gamma(n+1+\alpha/2)}.\]
The same result has been obtained in \cite[Prop.~1]{net:Zhang14twc} by calculating the limiting outage
probability $F_{\sir}(\theta)$ as $\theta\to 0$.

For $\alpha=4$, $\E(\isr^{(!n)})=\frac{2}{n+2}$, and
the asymptotic gain per \eqref{asym_gain} is simply
\[ G_{\rm silence[4]}=\frac{1}{\E(\isr^{(!n)})}=1+\frac n2 .\]

\begin{figure}
\centerline{\epsfig{file=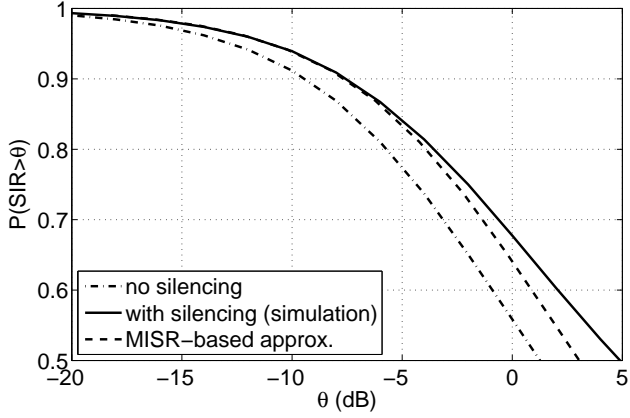,width=\figwidth}} 
\caption{Gain from silencing one base station in HIP model for $\alpha=4$.}
\label{fig:bs_silence}
\end{figure}

\figref{fig:bs_silence} shows the SIR distributions for the HIP model without silencing,
for the HIP model with silencing of one BS, and the MISR-based approximation.
The approximation is tight for success probabilities above $3/4$; after that, it is
pessimistic.

\subsection{Base station cooperation for worst-case users}
\begin{figure}
\centerline{\epsfig{file=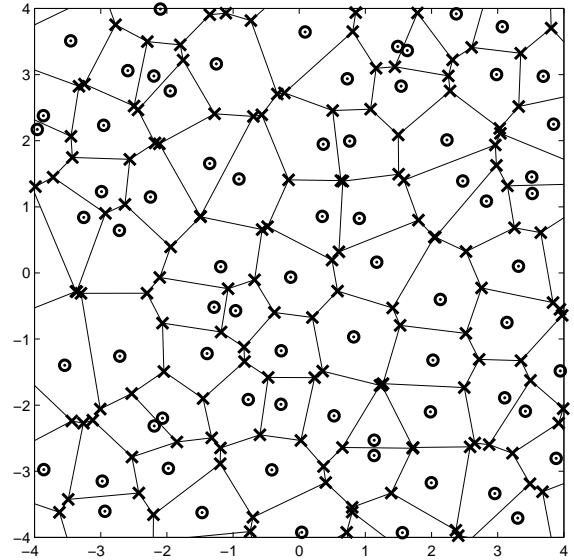,width=.9\figwidth}}
\caption{Illustration of worst-case user locations. Base stations are marked by $\odot$, and the crosses $\times$ are the vertices of the Voronoi tessellation and mark the locations of
the worst-case users. These users have the same distance to the 3 nearest BSs.}
\label{fig:voro_worst}
\end{figure}

We focus on {\em worst-case users} in the single-tier HIP model, which are the ones located at the vertices of the Voronoi
tessellation \cite{net:Jung13cl,net:Nigam14tcom}. These locations are marked by $\times$ in
\figref{fig:voro_worst}, and the SIR ccdf is denoted as $\bar F_{\sir}^\times$ accordingly.
Worst-case users are at a significant disadvantage if they are served by a single BS since
they have two other BSs at the same distance\footnote{In contrast to what is claimed in
\cite{net:Jung13cl}, these worst-case locations do {\em not} form a PPP.}.

With (non-coherent) joint transmission\footnote{The amplitudes of the three signals are adding up, and the combined received signal is still subject to Rayleigh fading.} from the 3 equidistant BSs and $\alpha=4$,
the ccdf follows from \cite[Thm.~2]{net:Nigam14tcom} as
\begin{equation}
   \bar F_{\sir[4]}^{{\times,{\rm coop}}}(\theta)=\bar F_{\sir[4]}^{{2}}(\theta/{3})=\Big(1+\sqrt{\theta/{3}}\arctan(\sqrt{\theta/{3}})\Big)^{-2}. 
   \label{coop4}
 \end{equation}
where $\bar F_{\sir[4]}$ is the SIR ccdf for the typical user in the HIP model given in \eqref{arctan}.
The factor of $3$ is due to the gain in signal power, while the 
exponent of 2 is due to the larger distance of the nearest BS than in the case of the
typical user. 

\begin{figure}
\centerline{\epsfig{file=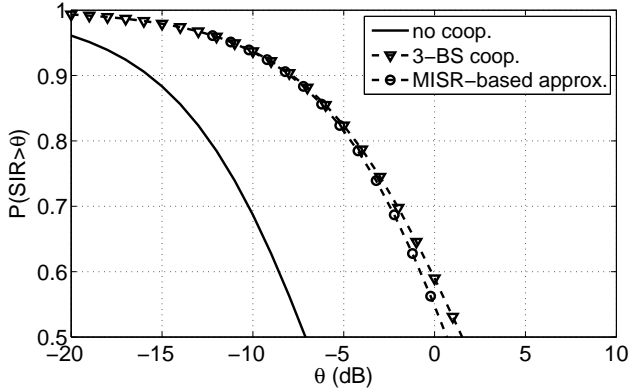,width=\figwidth}} 
\caption{SIR ccdf for worst-case users without cooperation and with 3-BS cooperation (from 
\eqref{coop4}) and MISR-based approximation for $\alpha=4$. Here $G=6$ (7.8 dB).}
\label{fig:worst}
\end{figure}

Without cooperation, $\misr^\times=2+4/(\alpha-2)$, and for
$n\in\{2,3\}$ BSs cooperating, it follows from \cite[Thm.~4]{net:Nigam14tcom} that
\[ \misr^\times_{n-{\rm coop}}=\frac{4+(3-n)(\alpha-2)}{n(\alpha-2)} .\]
So for $n=3$, the gain is
\[ G_{3-{\rm coop}}=\frac{{\rm MISR}^\times}{\rm MISR_{3-coop}^\times}=3+\frac32(\alpha-2) .\]
\figref{fig:worst} shows the
SIR distribution for worst-case users without cooperation, with cooperation from
the 3 nearest BSs, and the MISR-based approximation.

\subsection{Non-Poisson deployment}
An SIR gain can also be obtained by deploying the BSs more regularly (repulsively) than a PPP.
This gain has been termed {\em deployment gain} in
\cite{net:Guo13twc,net:Guo14tcom}. Exact closed-form results for the SIR distribution for
non-Poisson deployments are impossible to derive. However, the MISR-based 
approximation, relative to the HIP model, is fairly easy to evaluate and quite accurate.

\begin{figure}
\centerline{\epsfig{file=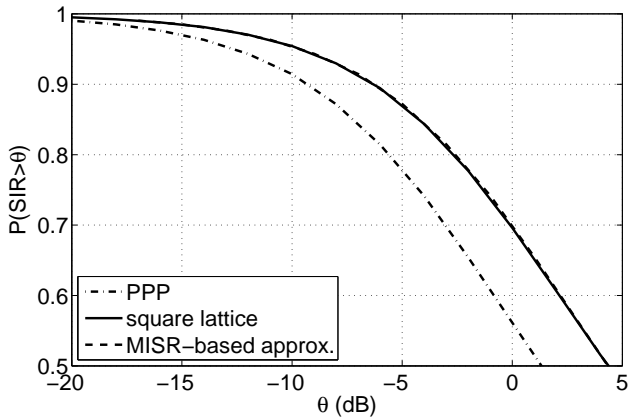,width=\figwidth}} 
\caption{SIR ccdf for the square lattice and MISR-based approximation for $\alpha=4$ (Rayleigh fading).}
\label{fig:square}
\end{figure}

Simulations show that the MISR of the square
lattice is quite exactly half of that of the PPP, irrespective of the path loss exponent, \ie,
the deployment gain is 3 dB.
\figref{fig:square} shows that the resulting approximation is accurate over a wide
range of $\theta$. As a result,
\[ \bar F_{\sir[4]}^{\rm sq}(\theta) \approx \bar F_{\sir[4]}(\theta/2) ,\]
where $\bar F_{\sir[4]}$ is given in \eqref{arctan}, is a very good analytical approximation
to the SIR distribution in the square lattice. For the triangular lattice (hexagonal cells),
the gain is slightly larger, about 3.4 dB, which is the maximum achievable.

\section{General fading and diversity}
\label{sec:naka}
So far we have discussed the case of Rayleigh fading. The MISR framework
easily extends to other types of fading or transmission schemes with diversity
(\eg, coherent BS cooperation, MIMO, retransmission). The diversity under interference (DUI) is defined as \cite{net:Haenggi13twc,net:Zhang14twc}

\[ d\triangleq\lim_{\theta\to 0} \frac{\log F_{\sir}(\theta)}{\log\theta} .\]

When comparing SIR
distributions, the two schemes need to provide the same diversity gain $d$ (\ie, have the same
asymptotic slope in the outage curve when plotted on a log-log scale),
otherwise the asymptotic gain may be undefined or not helpful in approximating one ccdf
with the other.

\begin{figure}
\centerline{\epsfig{file=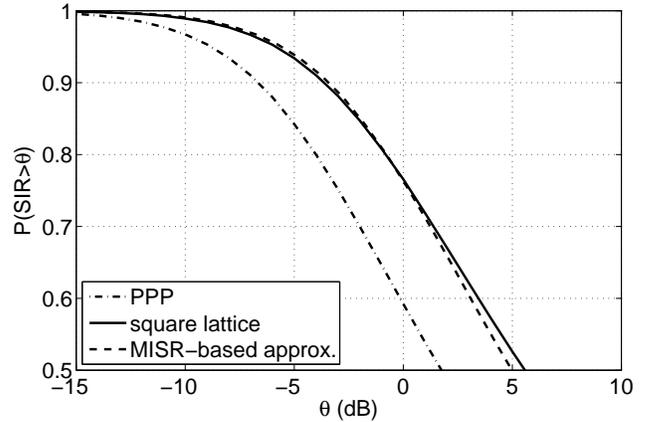,width=\figwidth}} 
\caption{SIR ccdf of square lattice and MISR-based approximation for $\alpha=4$ and Nakagami-2 fading.}
\label{fig:square_naka2}
\end{figure}

For example, if the fading distribution satisfies $F_h(x)\sim a x^m$, $x\to 0$, (as, \eg, in Nakagami-$m$ fading)
\[ F_{\sir}(\theta)\sim a \theta^m \E(\isr^m), \]
the diversity order is $m$---if the $m$-th moment of the $\isr$ is finite.\footnote{%
For the PPP, it can be shown that all moments of the $\isr$ are finite. Whether it holds
for {\em all} stationary point processes is under investigation.}
The asymptotic gain follows as
\[ G^{(m)}= \left(\frac{\E(\isr_1^m)}{\E(\isr_2^m)}\right)^{1/m}
\approx G^{(1)},\]
where the approximation by $G^{(1)}$ holds since the factor $\E(\isr^m)^{1/m}/\misr$ is
about the same for both schemes and thus cancels approximately. This is illustrated in
\figref{fig:square_naka2} for Nakagami-2 fading and the square lattice. The shift by
$G^{(1)}=3$ dB still yields a very good approximation.

\section{Conclusions}
The SIR distributions of two transmission schemes or deployments in cellular networks that provide
the same diversity gain are, asymptotically, horizontally shifted version of each other, and the asymptotic
gap between them is quantified by the ratio of the MISRs of the two schemes. We demonstrated 
that this asymptotic gap also provides a good approximation for the gap at finite $\theta$.

If the MISR cannot be calculated analytically, it is relatively easy to determine by simulation since
it only depends on the BS and user locations and the transmit power levels, but not on the fading.
Due to its tractability, the HIP model is the prime candidate as a baseline model against which
other schemes can be evaluated. So even if it is not be accurate in all situations, it serves the
important role of a reference model.

We anticipate that future networks will not be based on a strict cellular architecture
but will become {\em amorphous} due to cooperation between BSs at different levels, relays,
and distributed antenna systems. Since an exact analytical evaluation for the SIR distribution
for these sophisticated and cognitive architectures seems hopeless, we believe that the proposed
MISR framework will  play an important role in the analysis of such emerging  {\em amorphous networks}.
\vspace*{-1mm}

 \bibliographystyle{IEEEtr}

 \end{document}